\documentclass[sigconf, authorversion, nonacm]{acmart}
\usepackage[utf8]{inputenc}

\usepackage{tikz}
\usepackage{subfigure}
\usepackage{enumitem}
\usepackage{multirow}
\usepackage{cleveref}
\usepackage{bbding}
\usepackage{pifont}
\usepackage{wasysym}
\newcommand{\uscode}{code comprehension}
\newcommand{\ucode}{code comprehension }

\newcommand{\UCode}{Code Comprehension}
\newcommand{\Eye}{\emph{Eye}}
\newcommand{\Eyes}{\emph{Eye }}
\newcommand{\beginners}{beginners}
\newcommand{\acite}[1]{\citeauthor{#1}~\cite{#1}}
\newcommand{\pvalue}{$p$-value}
\newcommand{\pa}{\emph{Prob1}}
\newcommand{\pb}{\emph{Prob2}}
\newcommand{\module}[1]{module~#1}
\newcommand{\Module}[1]{Module~#1}
\newcommand{\xmark}{\ding{55}}%

\begin{document}
\title{Eye: Program Visualizer for CS2}

\author{Aman Bansal}
\email{aman0456b@gmail.com}
\orcid{0000-0001-7771-5459}
\affiliation{%
  \institution{Indian Institute of Technology Bombay}
}
\author{Preey Shah}
\email{preeyshah@gmail.com}
\orcid{0000-0001-7771-5459}
\affiliation{%
  \institution{Indian Institute of Technology Bombay}
}
\author{Sahil Shah}
\email{sahilshah00199@gmail.com}
\affiliation{%
  \institution{Indian Institute of Technology Bombay}
}


\begin{abstract}
  In recent years, programming has witnessed a shift towards using standard libraries as a black box.
  However, there has not been a synchronous development of tools that can help demonstrate the working of such libraries in general programs, which poses an impediment to improved learning outcomes and makes debugging exasperating.
  In this paper, we present a tool \Eyes, which is an \textit{interactive visual interpreter} that provides an intuitive representation of code execution and commonly used data structures in the C++ STL library.
  Eye provides a comprehensive overview at each stage during run time including the execution stack and the state of data structures.
  The modular implementation allows for extension to other languages and modification of the graphics as desired.
  
  \Eyes opens up a gateway for CS2 students to more easily understand myriads of programs that are available on online programming websites, lowering the barrier towards self-learning of coding.
  It expands the scope of visualizing data structures from standard algorithms to general cases, benefiting both teachers as well as programmers who face issues in debugging.
  The interpreting nature of Eye also provides space for a visualizer that can describe the execution and not only the current state.
  We also conduct experiments to evaluate the efficacy of \Eyes for debugging and comprehending a completely new code.
  Our findings show that it becomes faster and less frustrating to debug certain problems using this tool, and also makes understanding a new code a much more pleasant experience.
\end{abstract}




\maketitle
\section{Introduction}
With the increasing popularity of Computer Science (CS), the number of students interested in a formal CS education is ever-growing and thus is growing the need for CS instructors to move from a standard write-on-board teaching style to a more productive methodology.
The advent of CoViD-19 and social distancing has globally amplified this demand by disadvantaging the conventional methods.
Instructors now need to both effectively teach over video conference and empower students to continue learning on their own without direct support from the instructor or the TAs.
Satisfying this demand requires access to tools that can facilitate self-learning and allow students to further expand their skill set by making use of existing programming resources (such as online programming websites). Appropriate tools that would help understand standard libraries and the relevant algorithms would go a long way in furthering this goal.

In this paper, we restrict ourselves to the education of students who satisfy the following criteria: (i) Have sufficient introductory programming knowledge (`CS1' curriculum equivalent), (ii) Learning data structures and algorithms  (`CS2' curriculum equivalent), and (iii) Practicing programming problems to hone programming skills. We refer to these students as \emph{\beginners} in the paper. We now enumerate the specific difficulties that challenge these \beginners:
\begin{figure*}[!ht]
    \centering
    \fbox{\includegraphics[width=0.92\textwidth]{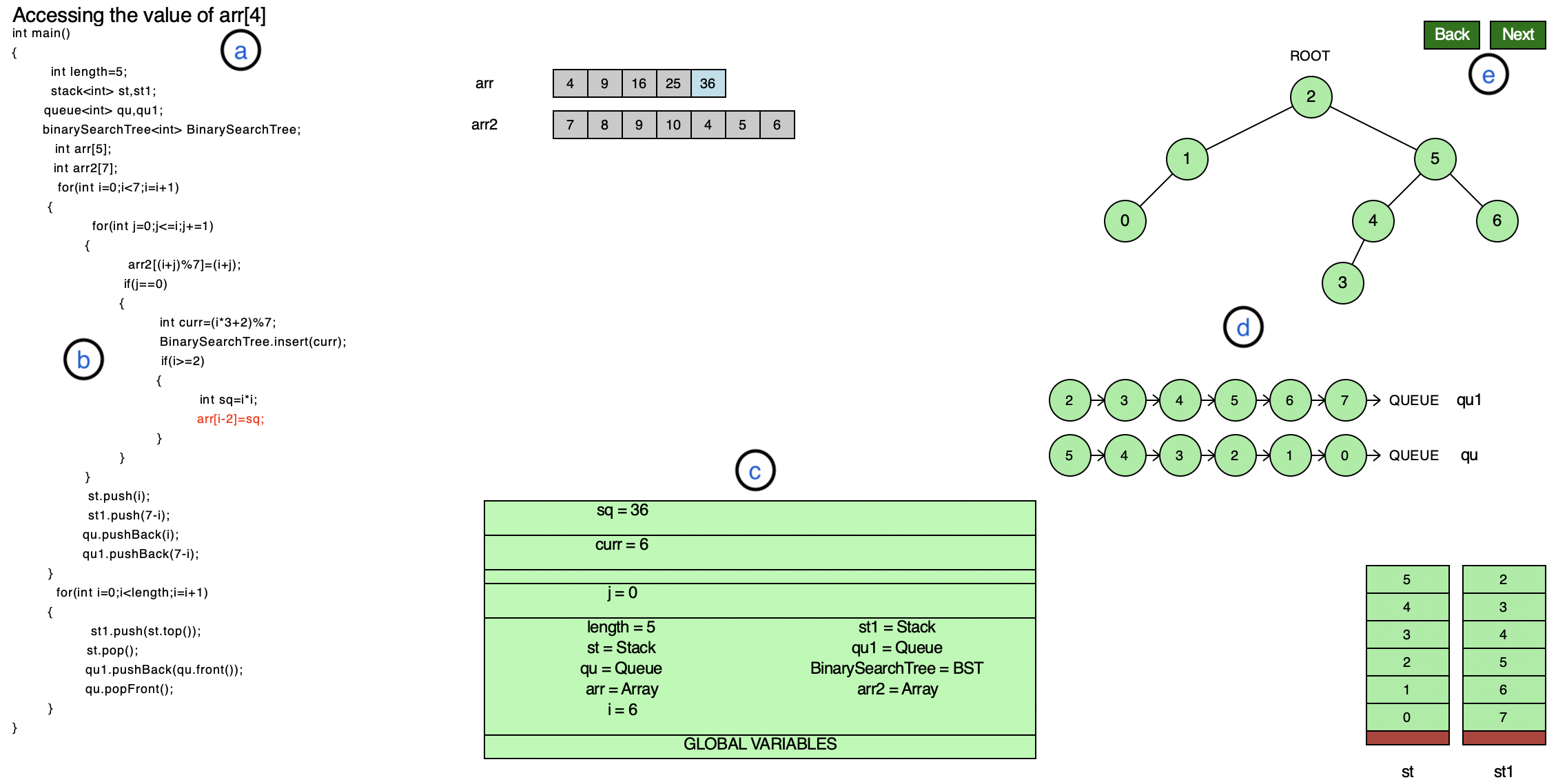}}
    \caption{Main Screen of \Eye. (a) Explanation of code being executed. (b) Program with code snippet being executed in red. The names of data structure classes are kept intuitive for better understanding. (c) Variables in the execution stack, with a different block for each scope. (d) Visualization of data structures like queue, stack, binary search tree, and array. (e) Interactive buttons to allow the users to move back and forward.}
    \Description{The main screen of Eye. It shows how the current display looks and positions elements.}
    \label{fig:main}
\end{figure*}
\begin{enumerate}
    \item The primary challenge they face is understanding the working of the well-known data structures and learning the different algorithms that manipulate them.
    While there are standard algorithms that demonstrate the usage of such structures, there is no tool that visualizes data structures in an arbitrary program.
    Henceforth, we refer to this problem as \emph{learning}.
    \item The second challenge they face is that while practicing programming problems, much to their dismay, an inordinate amount of their time is spent
    debugging, which is a very frustrating process \cite{debugtime, debugfrustrate}.
    In fact, it is considered by many as the most difficult part of learning programming \cite{debugStudy}.
    Compile-time or run time errors have some helpful message or stack trace which can be used intelligently to simplify debugging \cite{mercuri, debugML}, but errors that cause the program to give wrong results without obstructing it are much harder to find and fix.
    This problem is accentuated for bugs resulting from an incorrect understanding or usage of standard libraries and their algorithms.
    We call these bugs \emph{logical bugs} and we refer to this problem as \emph{logical debugging}.
    \item Moreover, the \beginners\ would inevitably be unable to solve some problems, engendering a need for explanatory solutions.
    Most of the time, these solutions do not exist and the most readily, sometimes the only, available option is to find the working code of the problem setter (or someone else) and understand it.
    This is markedly more pronounced for problems on online programming websites.
    Here they face their third challenge - understanding a completely new code.
    We refer to this problem as \emph{\uscode}.
    Notably, most of these problems require the usage of standard data structures, whose solutions are written by other peers in varying programming styles.
\end{enumerate}

Our contributions toward mitigation of these problems are:
\begin{enumerate}
    \item We introduce \Eye, an interactive pedagogical tool that visualizes a program's execution as it runs. It demonstrates properties and usage of data structures in a general environment, thereby helping in learning, logical debugging, and \uscode.
    
    \item We present two experiments, along with their methodology and results, which analyze the efficacy of \Eyes for logical debugging and \uscode.
    The first experiment measures the benefit of using \Eyes in debugging programs of which the subjects have some high-level knowledge, including the algorithm, the role of variables, and the loop invariants.
    The second experiment measures the benefit of using \Eyes in understanding a program on a high-level (such as time complexity, loop invariants, and role of data structures) given the problem statement and the program.
\end{enumerate}

We now specify some important properties and reason that they are essential for widespread adoption of any visualization tool. All these properties are satisfied by \Eye.

\begin{enumerate}[label=P\arabic*]
    \item \textbf{Completeness}: \label{pcomplete} It should support CS2-equivalent courses by visualizing data structures classes (such as C++ STL). This is required for covering the CS curriculum of \beginners.
    \item \textbf{Flexibility}: \label{pflex} It should provide flexibility to change the display (beyond CSS based changes) so that the instructor can modify it without much trouble.
    This is needed because instructors would want to focus on different aspects of the program in different lectures and would require changes like zooming in on a data structure, adding animations, or using some external display library of their choice.
    We believe that the lack of this flexibility can drastically decrease adoption among different universities.
    Allowing these changes but with considerable modifications can also scare away instructors \cite{newsorva}.
    \item \textbf{Awareness}: \label{paware} It should allow the addition of `program-aware' features, including but not limited to explanatory text and variable scoping.
    This is needed because these features can significantly improve the understanding of the program. Tools that are dependent on execution trace cannot demonstrate such features.
    \item \textbf{Accessibility}: \label{paccess} It should be able to run and display the visual elements in a web browser.
    This is for ensuring that every user can use it from anywhere without any hassle of installation or compatibility.
    \item \textbf{Modularity}: \label{pmodel} It should support multiple languages or be modular enough to support new languages with minimal back-end changes.
    This ensures that the tool is customizable for different universities and instructors and requires that the language-specific part be separate from the remaining implementation.
    \item \textbf{Interactivity}: \label{pinteract} It should be interactive, allowing the \beginners\ to go back and forth as per their convenience.
    This is required because interactive tools make the students learn better than passive tools \cite{tanen,reco}.
\end{enumerate}
We show how \Eyes satisfies these properties in \Cref{sec:impl,sec:design}. In \Cref{sec:exp}, we present our results concerning the effect of \Eyes on debugging and \uscode.
We then conclude in \Cref{sec:conc}.
\section{Related Work}
\subsection{Learning}
The concept of using visualizations to avoid the drawbacks of on-the-board teaching (see \cite{orsega}) and improve the understanding of algorithms is not new.
Studies have shown that the amount of time students spend on interactive visualization tools correlates with their performance \cite{tutor,middle-long}.
Therefore, \Eye, as a visual tool, has much potential for improving learning.

In the past few decades, a large number of program visualization tools have been created. \acite{sorva} and \acite{newsorva} give an overview of many such tools.
However, when scrutinized closely, many of these lack our required properties.
We give a comparison with some of the prominent program visualization tools in \Cref{works}. 
We note that Jsvee (\Cref{works}) provides only limited flexibility, which is also discussed in \Cref{sec:impl}.
\begin{table}[H]
\caption{Comparison with Existing Visualization Tools}
\label{works}
\begin{tabular}{ccccccc}
    \toprule
    \textbf{Tools} & \textbf{\ref{pcomplete}} & \textbf{\ref{pflex}} & \textbf{\ref{paware}} & \textbf{\ref{paccess}} & \textbf{\ref{pmodel}} & \textbf{\ref{pinteract}}\\
    \midrule
    Jeliot 3 \cite{jeliot} & \xmark & \xmark & \xmark & \xmark & \xmark & \checkmark\\
    jGRASP \cite{jgrasp} & \checkmark & \xmark & \xmark & \xmark & \xmark & \xmark\\
    Jsvee \cite{jsvee} & \xmark & \checkmark & \checkmark & \checkmark & \checkmark & \checkmark\\
    Python Tutor \cite{tutor} & \xmark & \xmark & \xmark & \checkmark & \checkmark & \checkmark\\
    UUhistle \cite{uuhistle} & \xmark & \xmark & \checkmark & \xmark & \xmark & \checkmark\\
    ViLLE \cite{ville} & \xmark & \xmark & \checkmark & \xmark & \checkmark & \checkmark\\
    \hline
    \Eye & \checkmark & \checkmark & \checkmark & \checkmark & \checkmark & \checkmark\\
    \bottomrule
\end{tabular}
\end{table}
\subsection{Logical Debugging}
\acite{syndebug1} have shown that debugging requires skills distinct from general programming skills.
Yet, these skills are not explicitly taught by the instructors and the students have to learn debugging techniques on their own \cite{debugk12}.
The industry debuggers do not help either as they are meant for professionals and tend to be too difficult to use and understand by \beginners.
Furthermore, most do not show the internals of a library.
As a result, they might not catch an incorrect update to a data structure until the program's end.
We seek to cover this gap with our tool.
In fact, we believe that \Eyes can be a crucial stepping stone for \beginners\ aiming to use industry debuggers. 

The general opinion in relevant literature is that the use of a tool for debugging is indeed beneficial.
\acite{debugexp} comment on how such tools should be used and integrated with the conventional teaching methods for better results.
\acite{debug1} discuss that introducing such tools earlier than later is even more beneficial.
One criticism of debugging tools is that they can help find a bug but cannot help correct it.
However, \acite{debugresults} report that beginners face the most difficulty finding the bug and that once found, fixing it does not take much effort.

We do not expect \Eyes to be a panacea, but these results are prompting enough to expect that it can help in debugging.
To our surprise, a literary survey to find a paper mentioning the effectiveness of such a tool for debugging purposes yielded no result.
Therefore to validate \Eye's potential, we devised and conducted our own experiment (see Section \ref{sec:exp}).
\subsection{\UCode}
This problem has been studied under the domain of \emph{algorithm visualization}~(AV), which is different from \emph{program visualization}~(PV).
The goal of AV is to visually aid the learning of an algorithm and not visualize a general program \cite{avpv}.
JSAV \cite{jsav} is one such prominent AV library for data structures. Our \ucode problem differs from this by focusing on \Eye, a PV tool.

This problem has also been studied indirectly in the debugging literature with the motivation of analyzing difficulties in debugging someone else's code and \acite{main} discuss this in their comprehensive literary survey on debugging.
\acite{gould} argue that students first spend time understanding the given code and only then start finding bugs. This separation has been further corroborated by other studies \cite{gouldski,syndebug1}.
Moreover, \acite{elsecode} provide strong evidence that the skills needed to understand the system are not necessarily connected to the skills needed to locate the error.
We have discussed the latter in the previous subsection. Regarding the former, we could not find a result showcasing the efficacy of a program visualization tool for \uscode, let alone with data structure libraries.
Therefore, we devise and conduct our own experiment (see Section \ref{sec:exp}).
\section{Design Overview}\label{sec:design}
\begin{figure}
        \centering
        \subfigure[The currently executing line is highlighted and explained.]{
            \includegraphics[width=0.2\textwidth]{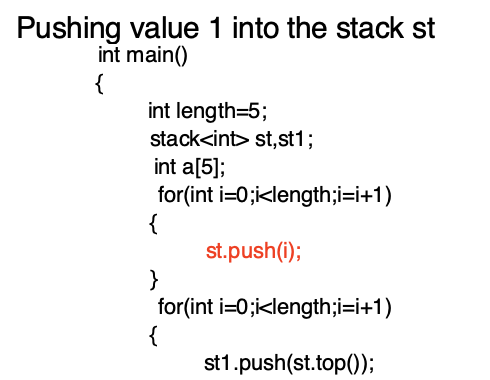}
            \label{fig:code}
        }
        \subfigure[The fourth element of the array is highlighted as it is being accessed.]{
            \includegraphics[width=0.2\textwidth]{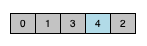}
            \label{fig:array}
        }
        \subfigure[The execution stack with scope separation as seen for variable \textit{length}. The empty region shows a scope where no variable was declared.]{
            \includegraphics[width=0.3\textwidth]{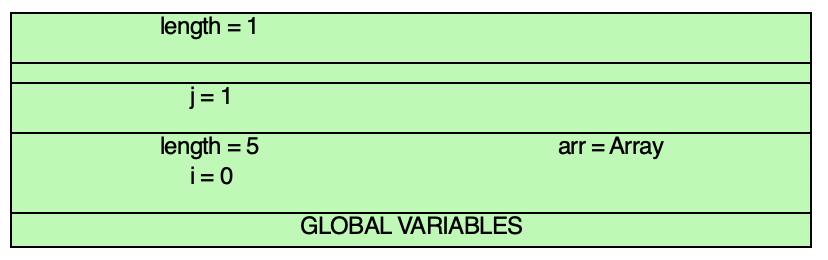}
            \label{fig:stack}
        }
        \subfigure[ Currently active function stack in a different color.]{
            \includegraphics[trim={0 0.75cm 0 0},clip,width=0.1\textwidth]{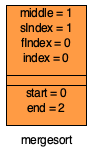}
            \label{fig:function}
        }
        \caption{Basic Design Features}
        \Description{The basic design features of \Eyes such as code highlighting, explanatory line, array highlighting, execution stack with scoping, differently colored execution stack for a function call.}
        \label{fig:line}
\end{figure}
\begin{figure}
        \centering
        \subfigure[Stacks]{
            \includegraphics[width=0.2\textwidth]{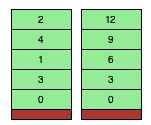}
            \label{fig:cstack}
        }
        \subfigure[Hash Table for integers with closed addressing and separate chaining for collision avoidance. The hash function used is modulo 6.]{
            \includegraphics[width=0.2\textwidth]{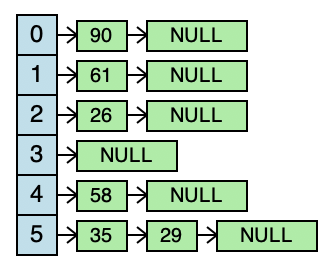}
            \label{fig:hash}
        }
        \subfigure[Queue]{
            \includegraphics[trim={0 1.5cm 0 0},clip,width=0.3\textwidth]{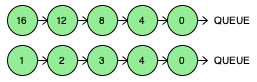}
            \label{fig:queue}
        }
        \caption{Data Structures }
        \Description{The visualization of various data structures such as stack, queue, and hash tables.}
        \label{fig:stl}
\end{figure}
In this section, we describe the functionality provided by our tool, including the essential elements that are common with different tools and some additional features which we believe are integral for our purpose.
\Cref{fig:main} shows the window with some fundamental elements on the canvas.
The tool currently supports \emph{C++} with \emph{STL}, but thanks to the modular design (see \Cref{sec:impl}), it can be easily extended to support other languages like \emph{Java} and \emph{Python}.
We now enumerate some of the basic elements common among other tools and detail how we supplement them to make them more descriptive.
\begin{itemize}
    \item An execution stack that shows all the variables and their current values. For clarity, data structures are shown outside the stack.
    We divide the stack into different sections to represent different scopes, as shown in \Cref{fig:stack}.
    Showing variables with different scopes in different sections of the stack elucidates the concept of scoping and expedites the detection of bugs due to \emph{variable shadowing}.
    Surprisingly, this simple feature was missing from other tools we studied.
    \item A new execution stack as soon as a new function starts executing. To make understanding easier, we color the currently active frame with a different color, as shown in \Cref{fig:function}.
    \item Besides displaying the source code with the current line highlighted, we provide an explanatory line summarizing the operation being executed (\Cref{fig:code}).
    It allows faster debugging by avoiding having to look at the syntactically dense code and reading the explanation instead for checking the correctness .
\end{itemize}

\begin{figure}
        \centering
        \subfigure[Root is being read.]{
            \includegraphics[width=0.215\textwidth]{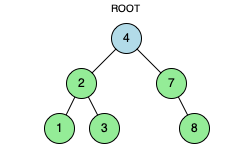}
        }
        \subfigure[Light Blue color signifies that the node is being read.]{
            \includegraphics[width=0.215\textwidth]{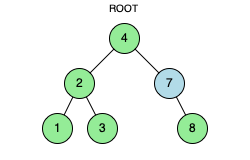}
        }
        \subfigure[Blue signifies that the node is being modified. Modifications include a change in child pointers or a change in value.]{
            \includegraphics[width=0.215\textwidth]{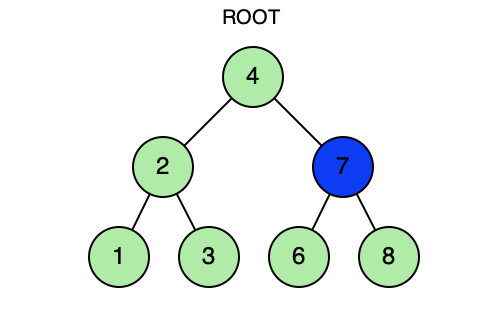}
        }
        \subfigure[Red signifies node deletion.]{
            \includegraphics[width=0.215\textwidth]{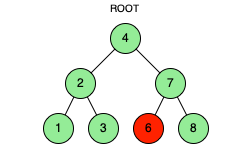}
        }
        \caption{Operations on Binary Search Tree (BST). (a), (b) and (c) show the insertion of value 6 into the binary search tree. (d) Deletion of node with value 6.} \Description{The coloring scheme used for explaining internal working of binary search tree. Different colors are used for reading, modifying and deleting.}
        \label{fig:color}
    \end{figure}

Now we enumerate some advanced design features of our tool.
\begin{itemize}
    \item Multiple data structures (STL constructs in C++) such as \text{vector} (array), \text{map} (binary search tree), \text{stack}, \text{queue}, \text{deque}, and \text{unordered$\_$map} (hash table) are supported (\Cref{fig:stl}).
    This lets beginners better grasp the working of these data structures and verify their state while debugging.
    
    \item Every access to these data structures is highlighted (including arrays, as shown in \Cref{fig:array,fig:main}).
    This speeds up the debugging process as students can skip the code or explanation and directly verify if all accesses (and assignments) are occurring as expected.
    For example, indexing errors, which are quite common among beginners, become noticeable due to this feature.
    It also helps in \ucode where the student can quickly see which value was read from or written onto a data structure.
    
    \item We carry this highlighting feature further to visually explain what happens internally in each data structure on a function call.
    For example, when an element is inserted or deleted in a binary search tree (\Cref{fig:color}).
    With this feature, we expect appreciable improvement in understanding when learning these data structures' working for the first time.
\end{itemize}

\section{Implementation Overview}\label{sec:impl}
\begin{figure*}
    \centering
    \begin{tikzpicture}[
roundnode/.style={circle, draw=green!60, fill=green!5, very thick, minimum size=7mm, text width=0.9cm, align=center},
squarednode/.style={rectangle, draw=red!60, fill=red!5, very thick, minimum size=10mm, align=center},
bigsquarednode1/.style={rectangle, draw=red!60, fill=red!5, very thick, minimum size=10mm,text width=2.1cm, align=center},
bigsquarednode2/.style={rectangle, draw=red!60, fill=red!5, very thick, minimum size=10mm,text width=2.6cm, align=center}
]
\usetikzlibrary{positioning}

\node[squarednode]      (code)                              {Source Code};
\node[roundnode]        (compiler)       [right = 0.5cm of code] {Module $\#$1};
\node[bigsquarednode1]      (ast)       [right = 0.5cm of compiler] {Canonical Code Representation};
\node[roundnode]        (pygraphics)       [right = 0.5cm of ast] {Module $\#$2};
\node[bigsquarednode2]      (json)       [right = 0.5cm of pygraphics] {Canonical Graphics Representation};
\node[roundnode]        (js)       [right = 0.5cm of json] {Module $\#$3};
\node[squarednode]      (display)       [right = 0.5cm of js] {Display};
\draw[green, very thick,-] (code.east) -- (compiler.west);
\draw[green, very thick,->] (compiler.east) -- (ast.west);
\draw[green, very thick,-] (ast.east) -- (pygraphics.west);
\draw[green, very thick,->] (pygraphics.east) -- (json.west);
\draw[green, very thick,-] (json.east) -- (js.west);
\draw[green, very thick,->] (js.east) -- (display);
\end{tikzpicture}
    \caption{Implementation Scheme}
    \Description{A flowchart which shows the implementation scheme for \Eye.}
    \label{fig:flow}
\end{figure*}
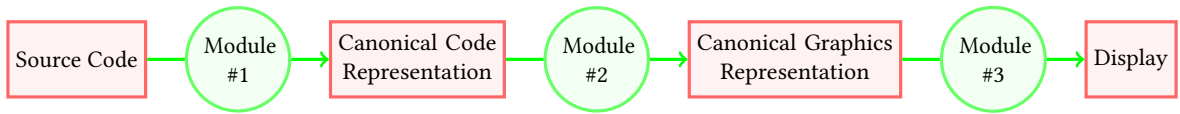
In this section, we give an overview of the implementation and show how \Eyes satisfies all the requirements that we assert are necessary for receiving wholesale traction.

The implementation is divided into three completely independent modules.
At a high-level, the role of these modules is summarized in \Cref{fig:flow}.
Before delving into these modules, we introduce the intermediate representations shown in the figure.\\\\
\textbf{Canonical Code Representation (CCR)}: It represents the source code in a format that is language independent.
It covers all the basic programming constructs usually taught in CS2, including data structures.
The primary benefit of introducing this representation is that adding support for different languages requires changes only in \module{1}, hence ensuring property \ref{pmodel} (Modularity).
The obvious choice for such a representation is an abstract syntax tree (AST).\\\\
\textbf{Canonical Graphics Representation (CGR)}: It is a representation of the information that \module{3} needs to create the graphics.
The reason for creating this intermediate stage is to ensure property \ref{pflex} (Flexibility).
Tools such as Jsvee \cite{jsvee} visualize the program parallelly with its execution.
This causes their visualization and execution semantics to get coupled, making it difficult to change the graphics.
Although it is possible to keep the coupling relaxed, it is natural to expect that an instructor would not be willing to understand the library's working to manipulate the visualization.
Another advantage is that an intermediate representation allows peeking into future frames to decide the display. For instance, if a data structure is not used in the next 50 frames, the instructor may reasonably wish to hide it for some frames.

CGR is created in the standard JSON format and includes, among other things, variables and the state of data structures.
It may appear similar to an execution trace, but our framework allows us to include significantly more information like the scope of variables and array accesses such as in \Cref{fig:array}.\\\\
\textbf{\Module{1}}: It converts the source code to an abstract syntax tree and is implemented in python.
We avoid using any external compiler as a black box because they impose extraneous restrictions and are usually daunting to modify for future developments.
The lexical analysis and parsing of the code were done using \emph{rply} library \cite{rply}.
The AST is made up of pre-defined python classes for every programming construct. Support for various C++ STL data structure libraries was added, ensuring property \ref{pcomplete} (Completeness).\\\\
\textbf{\Module{2}}: It converts the AST into a JSON object and is also implemented in python.
Every class in the AST implements an `exec' function which emulates its execution and generates the information required in CGR, including program-aware features, hence ensuring property \ref{paware} (Awareness).
To interpret and display data structures, we define custom classes with member functions which also create additional execution information.
For example, `insert' in a binary search tree can display each step of the algorithm if required, as shown in \Cref{fig:color}.
Enthusiastic instructors can modify these behaviors too, gaining more flexibility.

Currently, the whole CGR JSON object is returned in the end.
We can optionally pass it after every few line executions to reduce display latency in case of long execution times or infinite loops.\\\\
\textbf{\Module{3}}: It converts the CGR into an actual visual display.
It is implemented using HTML5, CSS, and JavaScript and can run on supported web browsers, ensuring property \ref{paccess} (Accessibility).
Buttons are present to go to the next or previous frame, ensuring property \ref{pinteract} (Interactivity).
Visualization can also be produced locally via \emph{graphics.py}, a basic graphic library of python~\cite{graphics.py}.
The current graphics can easily be further modified since our modular implementation allows the users great flexibility for this purpose.
They can pick colors, add animations, and even use external libraries to help them build appealing graphics.
\section{Experimental Results}
\label{sec:exp}
We design two experiments to measure the efficacy of \Eyes in debugging and \uscode. We try to answer the following research questions (RQ) via our experiments:\\\\
\textbf{RQ1(a)}: Does using \Eyes for debugging data structures based programs accelerate the debugging process?\\
\textbf{RQ1(b)}: Does using \Eyes for debugging data structures based programs reduce frustration usually seen in debugging process?\\
\textbf{RQ2(a)}: Does using \Eyes for understanding a new code improve the \ucode in a fixed amount of time?\\
\textbf{RQ2(b)}: Does using \Eyes for understanding a new code lead to better productivity in terms of time?\\

We contacted around 60 senior computer science undergraduates from our university, out of which 20 agreed to participate.
Before proceeding with the experiments, they were given a small demonstration and were asked to familiarize themselves with the tool.
The subjects ran the tool locally and not on the browser.
Each subject participated in two experiments for answering RQ1(a) and RQ2(a), and an anonymous survey to answer RQ1(b) and RQ2(b).

Due to social distancing, the experiments were conducted online using video conferencing software.
On average, each subject took around one hour to complete the experiment.
All the experiment material, including videos of some subjects taking the experiment, can be produced upon request.

\subsection{Experiment 1}
We conducted the experiment as follows:
\begin{enumerate}
    \item Subjects were given two problem statements (\pa\ and \pb) with buggy implementations of their solutions.
    The problems were based on data structures like \emph{stack} and \emph{queue}, and involved algorithms taught as part of CS2 curriculum (and hence were known to subjects).
    There was exactly one logical bug in both the implementations, and the subjects were asked to fix them.
    The time taken by the subjects to debug each program was recorded.
    \item The experiment was counterbalanced with respect to tool usage.
    Half of the subjects did \pa\ with \Eyes and \pb\ without (Group~1), and the other half did the opposite (Group~2).
    Subjects were assigned to these groups randomly.
    The problems were always given in the same order.
    Subjects using \Eyes were disallowed to edit or even see the code in any other application to ensure that they use \Eyes to debug.
    \item In a few cases, subjects could not debug the problem and gave up.
    The time for such subjects was then set to a default value larger than the time taken by any successful subject.
    \item Running the tool locally required a library installation that three people refused to do.
    Such subjects were allowed to debug both problems without \Eye.
    To somewhat offset the increase in number of without \Eyes measurements, one subject was asked to solve both the problems with \Eye.
\end{enumerate}

Group~1 had an average debug time of 1071.25 seconds for \pa\ and 1022.90 seconds for \pb\ while Group~2 had an average debug time of 778.75 seconds for \pa\ and 518.6 seconds for \pb. 
Due to random allocation, Group~1 had subjects with better debugging skills than Group~2 on average which is ratified by the average times of two groups - Group~1 took far more time that Group~2 for each of the questions.
To account for biases introduced by difference in debugging skills, we calculate the percentage of total debug time the subjects spent on the \pa\ (or equivalently \pb).
We consider these percentages to be random variables and test against the null hypothesis that the variables have the same mean for the two groups.
We had to eliminate the four subjects who did not have alternating tool usage for the two problems.
The average values for this measure is shown in \Cref{table:numbers}. The \pvalue\ for our data is $0.0578$.
These results show that \Eyes improves debug time.

\subsection{Experiment 2}
We conducted the experiment as follows:
\begin{enumerate}
    \item Subjects were divided into two equal-sized groups, randomly and independent of the previous experiment. One group used \Eyes for the experiment while the other had no restrictions.
    \item Both the groups were given a problem statement and a correct implementation of its solution.
    The solution was based on the \emph{deque} data structure of C++ STL and involved an algorithm new to the subjects.
    \item The subjects were first given 6 minutes to see the visualization (or go through the code) and try to understand how the algorithm is working. They were then given a link to a Google form which contained various questions. They were given 10 minutes to answer the quiz and were allowed to go back to the visualizer or the code during the quiz.
\end{enumerate}

We use the quiz score as a proxy for understanding.
Our null hypothesis for RQ2(a) was that there would be no considerable difference in the scores of the two groups.
We report the average percentage score for each group in \Cref{table:numbers}.
Although the group with \Eyes performed better, the difference was not statistically significant ($p=0.446$).
Nonetheless, given the biases against \Eyes (\Cref{sec:bias}) and the subjects' overwhelmingly positive opinion (\Cref{sec:opp}), we can reasonably expect that consistent use of \Eyes will show positive results.
On a hopeful note, \acite{middle-long} have shown that performance improvements do manifest when the students become conversant with a tool.


\subsection{Anonymous Survey} \label{sec:opp}
After the subjects had completed both the experiments, we asked them to fill an anonymous survey which contained two questions corresponding to RQ1(b) and RQ2(b).
The subjects were advised that this survey is for estimating the benefits of the tool so they should not bias their answer based on their particular experience in the experiment. The questions and their responses were as follows:

\begin{enumerate}[label=Q\arabic*(b)]
    \item \textbf{Question}: Assuming same time spent on debugging with \Eyes and without \Eye, how much do you think debugging with \Eyes can help in reducing frustration?\\
    \textbf{Options}: Ranging from 0\% to 100\%.\\
    \textbf{Response}: \Eyes reduces frustration by $61.43\%$ on average.\\
    \textbf{Conclusion}: A visualizer for libraries makes debugging remarkably less frustrating, thereby holding on a student's interest in programming for a longer time and raising enthusiasm for self-learning.
    
    \item \textbf{Question}: Assuming a time bound on your practice sessions and assuming you only want to understand someone else's submission of every problem you practice, how much can \Eyes improve your productivity (number of problems solved)?\\
    \textbf{Options}: Ranging from same to double productivity.\\
    \textbf{Response}: \Eyes can increase the number of problems solved by a factor of roughly $1.56$ on average.\\
    \textbf{Conclusion}: It demonstrates that students consider \Eyes useful for \ucode in that it allows faster understanding of someone else's code, greatly improving the utility of online programming websites.
\end{enumerate}
\begin{table}
\caption {Summary of the results of both experiments: With \Eye, there is a reduction in the average fraction of time taken to debug and an increment in average score, showing improvements in both use cases.}
 \label{table:numbers}
 \begin{tabular}{c|cc|c} 
 \toprule
   & \multicolumn{2}{c|}{Exp1 (Avg \% time)}& Exp2 (Avg \% score)  \\
    & \pa & \pb & \\
 \midrule
 Without \Eye & 58.3 & 51.4 & 59.2  \\
 With \Eye &  48.6 & 41.7 & 60.6 \\
 \hline
  \pvalue & \multicolumn{2}{c|}{0.0578} & 0.446\\
  \bottomrule
\end{tabular}
\end{table}
\subsection{Experimental Biases} \label{sec:bias}
\begin{itemize}
    \item Subjects were disallowed to use regular debugging techniques with \Eye. This potentially hampered their ability and added a bias against \Eye.
    \item Informal discussion with subjects after the experiment confirmed our suspicion of familiarity bias against \Eye.
    Many students primarily focused on the code (\Cref{fig:code}).
    Features like access highlighting (\Cref{fig:array}) were intended to reduce dependence on code-reading but were largely ignored.
\end{itemize}
\section{Conclusion}\label{sec:conc}
Data Structure libraries are widely used in schools, universities and industries for programming.
Visualization for such libraries is the need of the hour.
In this paper, we presented a tool \Eye, that offered a visual display of inner working of such libraries, thereby helping in learning, debugging and \uscode.
The efficacy of the tool was also tested through an assessment that showed encouraging results with the positive responses to the survey reinforcing its utility in practice.
Its design and functionality satisfy the properties that are required for widespread traction.

We believe that \Eyes will be extremely useful in universities and online courses for teaching purposes, and tweaks can be easily made to suit each course's requirements.
We plan to do a formal study to evaluate its efficacy in teaching when schools reopen and classes start.
In addition, extensive deployment on online programming websites can be done through integration with their IDEs (Integrated Development Environment) which requires a change in the display module.
Visualizations for more data structures and libraries in languages like Python and Java will lead to a greater adoption and expand its use cases to students learning other programming languages as well. 
\bibliography{ref.bib}
\bibliographystyle{ACM-Reference-Format}

\end{document}